\documentclass{webofc}
\usepackage[varg]{txfonts}   
\begin{document}
\title{Fission in a microscopic framework: from basic science to support for applications}
%
%

\author{\firstname{Ionel} \lastname{Stetcu}\inst{1}\fnsep\thanks{\email{stetcu@lanl.gov}} \and
        \firstname{Aurel} \lastname{Bulgac}\inst{2} \and
        \firstname{Shi} \lastname{Jin}\inst{2} \and
        \firstname{Kenneth J.} \lastname{Roche}\inst{2,3}\and
        \firstname{Nicolas} \lastname{Schunck}\inst{4}
}

\institute{Los Alamos National Laboratory, Los Alamos, New Mexico, 87545, USA
\and
           University of Washington, Seattle, Washington 98195-1560, USA
\and
           Pacific Northwest National Laboratory, Richland, WA 99352, USA
\and
           Lawrence Livermore National Laboratory, Livermore,  California 94551, USA
          }

\abstract{%
  Recent developments, both in theoretical modeling and computational power, have allowed us to make progress on a goal not fully achieved yet in nuclear theory: a microscopic theory of nuclear fission. Even if the complete microscopic description remains a computationally demanding task, the information that can be provided by current calculations can be extremely useful to guide and constrain more phenomenological approaches, which are simpler to implement. First, a microscopic model that describes the real-time dynamics of the fissioning system can justify or rule out some of the approximations. Second, the microscopic approach can be used to obtain trends, e.g., with increasing excitation energy of the fissioning system, or even to compute observables that cannot be otherwise calculated in phenomenological approaches or that can be hindered by the limitations of the method. We briefly present in this contribution the time-dependent superfluid local density approximation (TDSLDA) approach to nuclear fission, approach that has become a very successful theoretical model in many areas of many-body research. The TDSLDA incorporates the effects of the continuum, the dynamics of the pairing field, and the numerical solution is implemented with controlled approximations and negligible numerical corrections. The main part of the current contribution will be dedicated to discussing the method, and recent results concerning the fission dynamics. In addition, we present results on the excitation energy sharing between the fragments, which are in agreement with a qualitative conclusions extracted from a limited number of experimental measurements of properties of prompt neutrons.
}
\maketitle
%

\section{Overview and motivation}
\label{intro}

Fission is a complex many-body process that involves the shape evolution of a nuclear system from a compact object into two (or more) fission fragments. Phenomenological approaches to the description of fragment mass and charge distribution have reported a very good agreement with available experimental data \cite{fissionYields_th,PhysRevC.84.034613,PhysRevC.90.054609,PhysRevC.96.034603,PhysRevC.96.064616,PhysRevC.100.024612}.  In recent years, a host of developments, both in theoretical modeling and computational power, have allowed progress towards a more microscopic theory of nuclear fission \cite{Schunck_2016,SiSc16,younes2013,Bulgac:2015a,Regnier:2016xxt,PhysRevC.71.024316,PhysRevC.93.011304,Scamps:2018ab,PhysRevC.100.014615,PhysRevC.100.034615}. 

An \textit{ab initio} approach to the fission process, which starts from the inter-neutron interaction is well beyond not only short-term, but also long-term capabilities. Therefore, the method of choice for such processes is the density functional theory with different approximations for superfluid systems like Bardeen-Cooper-Schrieffer (BCS)  or superfluid local density approximation (SLDA)  \cite{Bulgac:2002uq,Bulgac:2007a}. Our group has used extensively the time-dependent superfluid approximation (TDSLDA) \cite{Bulgac:2013b}, which is particularly efficient to implement on today's supercomputers that use graphic accelerators. This approach has been has been very successful in describing a large number of phenomena and properties such as the ground state energies of quantum systems, pairing gaps, collective modes, quantized vortices, \textit{etc.} \cite{Bulgac:2011c,Bulgac:2013b,Bulgac:2013d,Wlazlowski:2015a,PhysRevLett.120.253002}. In nuclear physics, in addition to fission studies, TDSLDA has been applied to studies of the excitation of collective modes in deformed open-shell nuclei \cite{Stetcu:2011}, relativistic Coulomb excitations  \cite{Stetcu:2014}, and collisions of heavy nuclei \cite{PhysRevLett.119.042501}.

The description of the fission process does not stop with the formation of the fission fragments and the prompt neutron and gamma emission: the daughter nuclei, which are highly excited and far from stability, continue to beta decay and emit delayed neutrons and gamma rays until they reach the stability region. Because such processes are much slower that the dynamics from saddle-to-scission, their treatment is decoupled from the formation of the fission fragments. However, the properties of the emerging fragments define the properties of prompt fission neutrons and gammas, which have been extensively measured. Thus, indirectly, these measurements can validate information about the fission fragments. The main caveat is that other approximations are involved in the description of prompt fission neutrons and gammas using phenomenological approaches \cite{Litaize2015,SCHMIDT2016107,FREYA2.0.2,Talou2018}.

An assumption used in all codes that comput the properties of prompt fission particles is the fact that the latter are emitted from the fully accelerated fission fragments, thus ignoring the possibility of neutron emission at scission or during acceleration. Is this a good approximation? Some phenomenological models predict that the fraction of prompt neutrons emitted at scission is significant and cite experimental evidence to support the claim \cite{CRPRC10,dsm,PLB15,CCC,SNspectr}. While these models are not universally accepted, a microscopic approach should be able to give an answer to such a question. The only type of models that allow for such investigations are TDSLDA and related time-dependent approaches because the fission dynamics, including any neutrons emitted at scission or during acceleration, can be observed in time. 

Even if the assumption that the neutrons are emitted from the fully accelerated fragments is approximately correct, one has to keep in mind that in experiments the fragment properties cannot be measured before the emission of prompt neutrons and gamma rays. Thus, all the information regarding the excitation or spin distributions in fission fragments can only be inferred from measuring the properties of prompt neutron and gamma rays. Modelers are thus forced to introduce corrections that involve assumptions regarding neutron emission in order to extract the mass of the post-neutron emission mass distributions \cite{PhysRevC.94.054604}. In this context, models that constrain quantities like the excitation energy sharing between fragments or their spin play a crucial role. And while models that do not allow for a full separation of the fission fragments have been advanced with promising results \cite{albertsson2018microscopic}, an approach that allows for the computation of the excitation energies in each fragment can significantly reduce the theoretical uncertainties. We have recently established~\cite{PhysRevC.100.034615} that the properties of the fission fragments are truly established only when the separation between the tips of the fission fragments is of the order of $4\ldots 5$ fm, which brings into question the validity of the phenomenological models used so far.

Even though the work and features of the fission process presented here mostly rely on the microscopic TDSLDA, the main message of this contribution is not necessarily to advocate toward one direction or the other, as both phenomenological and microscopic approaches have their advantages and disadvantages. Rather, we should envision a synergistic picture, in which the microscopic models can inform the more phenomenological ones on the validity of approximations involved, or provide additional information on trends and other quantities that could be challenging to obtain otherwise. And, more importantly, significant applications of the fission process extend from national security to energy generation, or to the origin of elements 
in astrophysics. Thus, the need for guidance from theoretical models is urgent, and the best approach would be to combine both phenomenological and microscopic approaches that deliver relevant information for those applications.

\section{Brief theoretical background}
\label{sec-1}

We follow the dynamics of the nuclear system from compact configurations to separated fragments by solving in coordinate space the well known (self-consistent) time-dependent Schr\"{o}dinger equation
\begin{equation}
i\hbar\frac{\partial}{\partial t}\Psi(\vec r,t)=H(\rho_p, \rho_n, \cdots)\Psi(\vec r,t),
\label{eq:tdslda}
\end{equation}
where $\Psi(\vec r,t)$ are the 4-component quasiparticle wave functions, for both protons and neutrons, and the $4\times 4$ Hamiltonian matrix operator $H$ depends upon the instantaneous neutron and proton densities (normal, anomalous, spin, and current densities).  $H$ is derived from a given nuclear energy density functional by taking a functional derivative with resect to the density matrix and has a block structure with particle-hole and particle-particle components. For details see Ref. \cite{PhysRevC.100.034615}. The many-body wavefunction is approximated by a single Slater determinant composed of all the quasiparticle wavefunctions. In current simulations we include all quasiparticles that correspond to positive eigenvalues of the Hamiltonian $H$ in the presense of constraints. During the entire evolution, the many-body solution remains a quasiparticle Slater determinant, even when the full separation of the fragments is achieved. Thus, in the spirit of Kohn and Sham, the only physical quantities that can be obtained are one-body observables, in particular densities and currents, which can be used to calculate the properties of the fission fragments.

In numerical calculations, we discretize Eq. (\ref{eq:tdslda}) on a rectangular lattice, characterized by $N_x$, $N_y$ and $N_z$ points on directions $x$, $y$ and $z$, respectively.   In order to capture the dynamics of the system until the two fragments are fully separated, we have employed boxes with dimensions $N_x=N_y=24$ and $N_z=48$, with lattice constant $l=1.25$ fm on all directions, which is equivalent with a momentum cutoff of about 500 MeV, sufficiently high to describe low-energy physics. However, calculations are in progress to reduce the lattice constant by 40\%, which in turn requires performing even larger-scale calculations in order to preserve physical dimensions of the rectangular box in which the properties of the two fragments can be calculated after considerable separation.

In order to evolve the nuclear system in time, we employ the fifth order numerical method Adams-Bashford-Milne \cite{hamming1986}. In this approach, the self-consistent Hamiltonian in Eq. (\ref{eq:tdslda}) is applied twice per time step, with errors of order $\mathcal{O}(\Delta t^6)$, where $\Delta t$ is the numerical integration step in time. The derivatives are efficiently calculated via Fourier transforms, using GPU accelerators. On systems with GPUs, the bottleneck is often the need to transfer large amounts of data between CPUs and GPUs. We minimize the amount of data exchanged, by only transferring the densities for reduction over CPUs using MPI calls, and recompute the potentials on the fly on each GPU. This ensures almost perfect weak and strong scaling properties.

 The initial states are taken beyond the outer barrier on the zero temperature potential energy surface defined by constraints on quadrupole and octupole moments, and evolved using Eq. (\ref{eq:tdslda}) until the system splits into two fragments with their centers of mass separated by 30 fm. During the entire evolution, the solution remains one single quasiparticle Slater determinant. The properties of the two fragments are calculated by dividing the simulation box in two, and using the one-body densities in each half. Each fragment's energy can be obtained using the same functional used to evolve the full system, while their mass and charge are obtain by simply integrating the total and proton densities respectively.
 
We can use as input in simulations any Skyrme-type density functional. In the pioneering work presented in Ref. \cite{Bulgac:2015a}, we have opted for the generic SLy4 functional, that is not particularly well suited for fission calculations. In our latest work \cite{PhysRevC.100.034615}, we have used two functionals that provide a better description of fission barriers. However, in general we obtain similar results, with the exception of the time from saddle to scission, which can exhibit strong dependence on the choice of the density functional.

TDSLDA requires in general more extensive computational resources than other approaches to fission. However, we perform routinely such calculations, as they are well within existing computer power. Our codes can efficiently run on leadership-class machines Summit and Sierra at Oak Ridge or Lawrence Livermore National Laboratories respectively,  Pizdaint (CSCS Switzerland), as well as on smaller machines like a GPU cluster at Los Alamos National Laboratory or Tsubame at Tokyo Institute of Techmology in Japan. 
 
\section{Selected results}

We have recently published a series of results of the TDSLDA simulations \cite{PhysRevC.100.034615}. In contrast with Ref. \cite{Bulgac:2015a}, where only a limited number of initial conditions and different pairing correlations for the same energy functional were investigated, in Ref. \cite{PhysRevC.100.034615} we  started with a larger number of initial conditions considering different points on the potential energy surface. The starting states were chosen to have a large spread in quadrupole deformation ($Q_{20}$) and mass asymmetry ($Q_{30}$), with similar excitation energies with respect to the ground state. After performing the TDSLDA simulation, we found that, independently of the functional used, the fission fragments emerge with similar properties (mass, charge, deformations, etc.).  In all calculations, including those reported in Ref. \cite{Bulgac:2015a}, TDSLDA is consistent with expectations that the light fragment emerges deformed, while the heavy fragment is closer to spherical shape with very weak or collapsed pairing field, as it is expected to be close to a closed shell configuration. We have also shown that immediately after scission, the fragments have more elongated shapes, which quickly relax. 

The relatively small effect of the initial state on the final properties of the fission fragments can be understood in terms of the strong one-body dissipation that is present in the model. As a consequence, in the dynamics from saddle-to-scission, the system behaves as a viscous fluid, with the trajectory following predominantly the steepest descent, and  (collective) flow energy that remains almost constant over the duration of the motion until scission. This has two consequences: (i) invalidates adiabatic approaches in which the collective potential energy is transformed into collective kinetic energy, and (ii) provides justification to the overdamped Brownian motion approach to fission yields pioneered by M\"oller and Randrup \cite{fissionYields_th,PhysRevC.84.034613}. We have further investigated dissipation by performing a theoretical experiment, in which we apply quadrupole kicks to the nuclear system, and observe the behavior of the flow energy \cite{PhysRevC.100.034615}. We have observed that due to the strong one-body dissipation, the collective kinetic energy transferred to the system is very quickly converted into internal excitation energy. 

\begin{figure}
\centering
\includegraphics[width=7.5cm,clip]{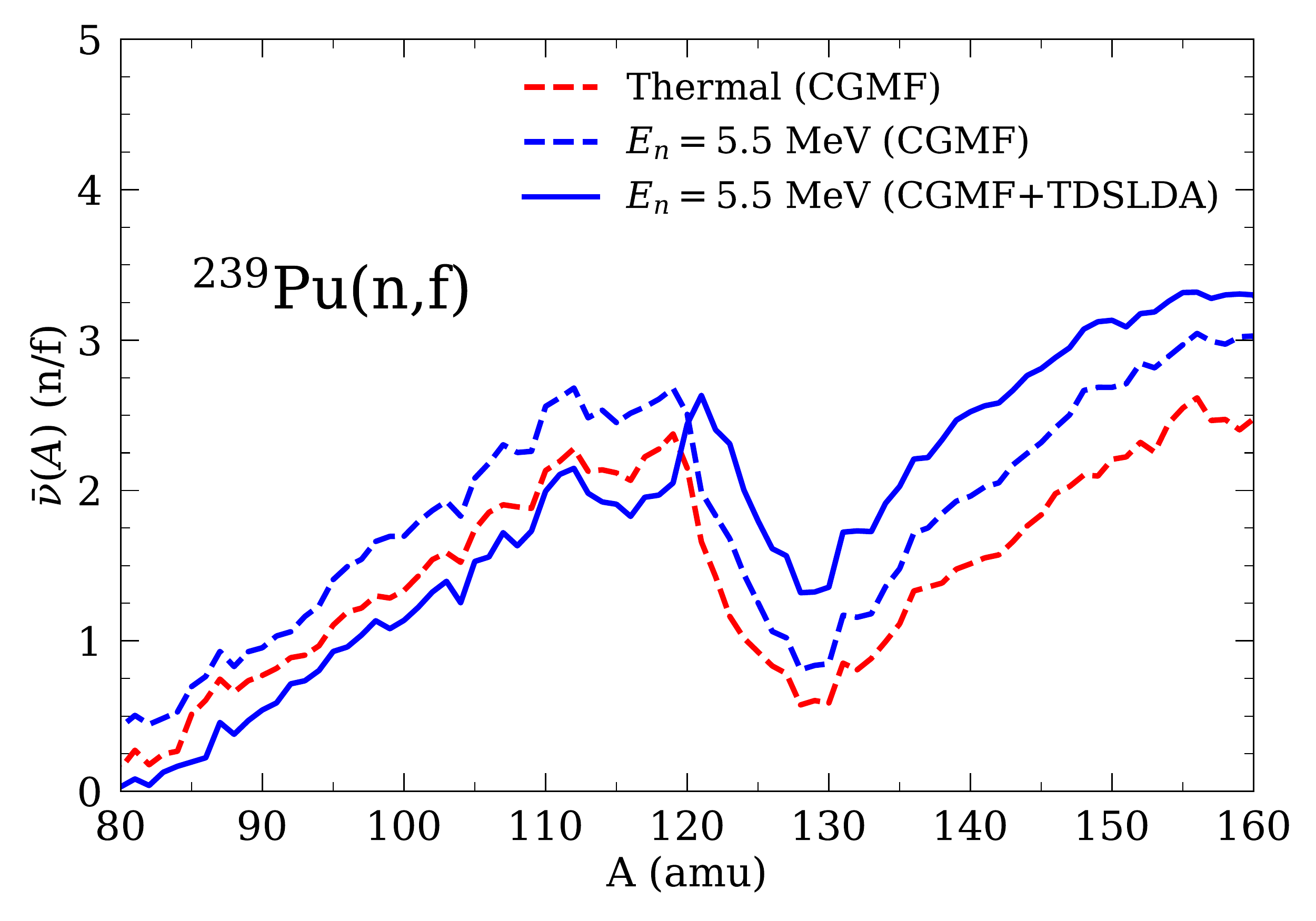}
\caption{The average prompt neutron multiplicity as a function of fragment mass, for two different incident energies, thermal and 5.5 MeV, for the $^{239}$Pu neutron-induced fission. With dashed lines we show the CGMF results in which no incident energy dependence is assumed, while with continuous line we show the 5.5 MeV CGMF results obtained using the parametrization of the incident neutron energy dependence extracted from TDSLDA calculations.}
\label{fig:nubarA}       
\end{figure}

We have used the results in Ref. \cite{PhysRevC.100.034615} to parametrize the excitation energy sharing between fragments as a function of the excitation energy in the fissioning system, which in turn can be expressed in a dependence of the ratio of temperatures of the incident neutron energy. In CGMF, the Los Alamos code that models the emission of prompt neutrons and gammas from the fission fragments, the energy sharing is parametrized by adjusting  the ratio of temperatures in the two fragments, $R_T(A)$ to available experimental data on the average neutron multiplicity as a function of pre-neutron emission fragment mass \cite{becker2013}. However, this kind of data is scarce, and mostly available for spontaneous fission or neutron induced fission with thermal neutrons. At the moment, the same $R_T(A)$ is used for all incident energies, which assumes the same qualitative sharing of the excitation energy between the fragments at all incident energies. However, experimental evidence \cite{PhysRevC.29.885,PhysRevC.34.218} suggest that with increasing the neutron incident energy, most of the extra excitation energy is transferred to the heavy fragment.  In oder to use the information from TDSLDA calculations, we have assumed a factorization of the ratio of temperatures as follows:

\begin{equation}
R_T(A,E_n)=R_T(A,E_\mathrm{thermal})\cdot f(E_n),
\label{eq:rtae}
\end{equation}
where $f(E_n)$ was assumed a simple function of the neutron incident energy $E_n$. From TDSLDA calculations, we have extracted the parameters of $f(E_n)$, and then applied an overall rescaling factor so that at thermal incident neutron energy $f(E_\mathrm{thermal})=1$. We impose this renormalization so that we reproduce the available thermal data on average prompt neutron multiplicity as a function of mass. The consequence of this parametrization is shown in Fig. \ref{fig:nubarA}, where we show that if no energy dependence is included in CGMF calculations, increasing the incident neutron energy produces an overall increase in the neutrons emitted by both the light and the heavy fragments (the two dashed curves). However, when Eq. (\ref{eq:rtae}) is used, the increase in prompt average neutron multiplicity is observed mostly for the heavy fragment, consistent with available experimental data on $^{235}$U and $^{237}$Np neutron-induced fission \cite{PhysRevC.29.885,PhysRevC.34.218}. The small decrease in the average multiplicity for the light fragment at 5.5 MeV vs. thermal incident energy is well within conceivable experimental uncertainties, and could also be a consequence of the simplified ratio of temperature treatment employed in CGMF.  Preliminary calculations using the parametrization extracted from $^{240}$Pu simulations have shown a good reproduction of the measured average prompt fission neutron multiplicity as a function of mass for $^{235}$U(n,f) and $^{237}$Np(n,f), which suggest that this parametrization is robust.

A solid theoretical model is required to parametrize the energy sharing between the two fragments above the threshold for multi-chance fission, where neutrons can be emitted before the remaining system undergoes fission. For incident energies above second-chance fission threshold, the prompt average neutron multiplicity as a function of fragment mass is not a well defined quantity because the neutron(s) emitted before fission cannot be identified as pertaining to a fragment or the other. Hence, even if the data on average multiplicities as a function of fragment mass existed, it would be hard to disentangle the contributions from different resulting from multi-chance fission and also challenging to parametrize the energy sharing mechanism for incident neutron energies above about 5.5 MeV. TDSLDA can, in turn, provide guidance and even quantitative parametrization for this mechanism.

\section{Outlook}

TDSLDA has proved already an important tool in understanding the fission process. It can provide support for some of the existing approaches to fission, and invalidate others. In this contribution, the discussion was limited to a few aspects of fission dynamics, including the parametrization of the energy sharing using information from TDSLDA. However, because it can follow the dynamics of the fissioning system until full separation, TDSLDA is ideally suited to provide information on all fission fragment properties, eventually as a function of the excitation energy of the fissioning nucleus. 

Current and future investigations can explore other important questions that influence the physics of prompt fission neutrons and gamma rays, which are some of the most readily available fission measurements. In particular, more work is necessary to study in more detail the neutrons that can potentially be emitted at scission, or whether part of the neutrons are emitted during acceleration. In addition, the initial spin distribution of the fission fragments can only be inferred from gamma observables, but its extraction is model-dependent \cite{IsomRatio2013}. TDSLDA output can be used to compute spin observables \cite{PhysRevC.99.034603,PhysRevC.100.034612}, and an investigation is undergoing.  

The need for reliable theoretical methods becomes even more clear when one considers that all the experimental measurements related to the properties of fission fragments are only possible after prompt neutron emission. Thus, primary fission fragment mass, total kinetic energy distributions or other quantities are always corrected for neutron emission, and thus is essential to understand whether neutrons are emitted at scission or during the acceleration of the fission fragments. Only models in which the dynamics of the system can be followed from before scission to well separated fragments can address such important questions.

\section*{Acknowledgments}

The work of I.S.  was supported by the US Department of Energy through the Los Alamos National Laboratory. Los Alamos National Laboratory is operated by Triad National Security, LLC, for the National Nuclear Security Administration of U.S. Department of Energy (Contract No. 89233218CNA000001).
The work of AB and SJ was supported by U.S. Department
of Energy, Office of Science, Grant No. DE-FG02-97ER41014
and in part by NNSA cooperative agreement DE-NA0003841.
The TDDFT calculations have been performed by SJ at the
OLCF Titan and Piz Daint and for generating initial configurations for direct input into the TDDFT code at OLCF Titan 
and NERSC Edison. This research used resources of the Oak
Ridge Leadership Computing Facility, which is a U.S. DOE
Office of Science User Facility supported under Contract No.
DE- AC05-00OR22725 and of the National Energy Research
Scientific computing Center, which is supported by the Office
of Science of the U.S. Department of Energy under Contract No.
DE-AC02-05CH11231. We acknowledge PRACE for awarding
us access to resource Piz Daint based at the Swiss National
Supercomputing Centre (CSCS), decision No. 2016153479.
The work of KJR is supported by US DOE Office of Advanced
Scientific Computing Research and was conducted at Pacific
Northwest National Laboratory and University of Washington.
The work of NS was supported by the Scientific Discovery
through Advanced Computing (SciDAC) program funded by
the U.S. Department of Energy, Office of Science, Advanced
Scientific Computing Research and Nuclear Physics, and it
was partly performed under the auspices of the US Department
of Energy by the Lawrence Livermore National Laboratory
under Contract DE-AC52-07NA27344. NS performed the
calculations of the initial configurations, of the ground state and
the finite-temperature properties of the FFs under the auspices
of the US Department of Energy by the Lawrence Livermore
National Laboratory under Contract DE-AC52-07NA27344.
Some of the calculations reported here have been performed
with computing support from the Lawrence Livermore National
Laboratory (LLNL) Institutional Computing Grand Challenge
program.

%
\bibliography{fissionBBL}
%
%

\end{document}